\documentclass[12pt]{article}

\usepackage{epsfig}

\input epsf
\begin{document}

\def\cA{{\cal A}}
\def\cL{{\cal L}}
\def\cD{{\cal D}}
\def\g{\gamma}
\def\ss{\scriptscriptstyle}
\def\gl{\gamma_{\scriptscriptstyle L}}
\def\gr{\gamma_{\scriptscriptstyle R}}
\def\gf{\gamma_5}
\def\d{\delta}
\def\e{\eta}
\def\diag{\hbox{diag}}
\def\pl{\partial}
\def\hf{{1\over 2}}
\def\ol#1{\overline{#1}}
\def\Dsl{\hbox{/\kern-.6700em\it D}} 
\def\dsl{\hbox{/\kern-.5300em$\partial$}}
\def\veps{\varepsilon}
\def\eps{\epsilon}
\def\ebar{\ol{\eta}}
\def\pbar{\ol{\psi}}
\def\zbar{\ol{\zeta}}
\def\xbar{\ol{\xi}}
\def\lbar{\ol{\lambda}}
\def\cbar{\ol{\chi}}
\def\eqa{\begin{eqnarray}}
\def\eeqa{\end{eqnarray}}
\def\eq{\begin{equation}}
\def\eeq{\end{equation}}
\def\la{\langle}
\def\ra{\rangle}
\def\foot{\footnote}
\renewcommand{\Im}{{\rm Im}\,}
\renewcommand{\Re}{{\rm Re}\,}
\newcommand{\beq}{\begin{equation}}
\newcommand{\bea}{\begin{eqnarray}}

\newcommand{\onefigure}[2]{\begin{figure}[htb]
\begin{center}\leavevmode\epsfbox{#1.eps}\end{center}
\caption{#2\label{#1}}
\end{figure}}

\def\nn{\nonumber}
\def\cc{\hbox{c.c.}}
\def\cL{{\cal L}}
\def\veps{\varepsilon}

\begin{titlepage}

\hfill hep-th/0407192

\hfill

\vspace{20pt}

\begin{center}
{\Large \textbf{POINCARE RECURRENCES AND TOPOLOGICAL DIVERSITY}}
\end{center}

\vspace{6pt}

\begin{center}
\textsl{M. Kleban $^{a}$, M. Porrati $^{b}$ and R. Rabad\'an $^{a}$}
\textsl{}
\vspace{20pt}

\textit{$^a$ School of Natural Sciences, Institute For Advanced Study\\ Olden Lane, Princeton, NJ 08540}

\vspace{4pt}

\textit{$^b$ Department of Physics, New York University\\ 4 Washington Pl.,
New York NY 10003}

\vspace{10pt}

\textit{}
\end{center}
\vspace{12pt}

\begin{abstract}

Finite entropy thermal systems undergo Poincar\'e recurrences.  In the
context of field theory, this implies that at finite temperature, timelike
two-point functions will be quasi-periodic.  In this note we
attempt to reproduce this behavior using the AdS/CFT correspondence by
studying the correlator of a massive scalar field
 in the bulk. We evaluate the correlator by summing
over all the $SL(2,Z)$ images of the BTZ spacetime.
We show that all the terms in this sum receive large corrections after at certain critical time, and that the result, even if convergent, is not quasi-periodic. We
present several arguments indicating that the periodicity will be very difficult
to recover without an
exact re-summation, and discuss several toy models which
illustrate this.  Finally, we consider the consequences for the information paradox.

\end{abstract}

\vspace{4pt} {\small \noindent
}
\vfill
\vskip 5.mm
 \hrule width 5.cm
\vskip 2.mm
{\small
\noindent e-mail: matthew@ias.edu, massimo.porrati@nyu.edu, rabadan@ias.edu
}
\end{titlepage}
\tableofcontents

\vspace{1cm}
\begin{verse}
Non domandarci la formula che mondi possa aprirti,\\
s\`{\i} qualche storta sillaba e secca come un ramo.\\
Codesto solo oggi possiamo dirti,\\
ci\`o che {\em non} siamo, ci\`o che {\em non} vogliamo.
\end{verse}

\section{Introduction}

In this
paper, we study the eternal BTZ \cite{BTZ} black hole in 2+1 dimensional AdS space.
As discussed in \cite{malda,bss,br,s}, the field theory dual to this black hole is thermal and generically has a discrete spectrum and finite entropy.  In such theories, timelike separated two point correlation functions should be quasi-periodic functions of time.  One can think of the correlator $\langle \phi(t) \phi(0) \rangle $ as the insertion of some
perturbation at time $t=0$, and then a measurement of the response at some later time $t>0$.  In thermal field theories, the short-time behavior is an exponential decay due to thermalization, but at long times the quantum version of the Poincar\'e recurrence theorem guarantees that the correlator will behave 
stochastically, returning (or at least coming arbitrarily close to) its 
initial value an infinite number of times.

The bulk dual to the short time exponential fall-off of the correlator is related to the fact that
black holes swallow anything that is thrown into them.  Classically, perturbed black holes ring with a quasi-normal frequency that has a non-zero imaginary part (so that the oscillation damps to zero exponentially), and semi-classical corrections {\it \'a la} Hawking simply create an exactly thermal atmosphere which presumably does not change the qualitative behavior.   Therefore, if field theory correlators are computed using free fields in a background
 BTZ spacetime, the time dependence is
a dying exponential, and the recurrences are gone.  This is closely related to the information
paradox of Hawking \cite{hawking}.  Restoring unitarity and resolving the information paradox, and matching with the CFT prediction, requires that the exact bulk correlator be quasi-periodic.

In this paper we study the behavior of the correlator in an ensemble of 2+1 dimensional bulk spacetimes dual to the boundary CFT. In the AdS/CFT correspondence, we are instructed to sum with an appropriate weight over all bulk geometries that fill a given boundary. This is just the Euclidean version of a
sometimes forgotten aspect of AdS/CFT: a CFT {\em
state} is dual to a quantum gravity {\em state}, not to a classical background
geometry. Such backgrounds can at best arise as saddle points in a field 
theory path integral.  One problem with this prescription is that, as far as 
we are aware, it does not arise from any well-defined bulk path integral, and 
among other problems this means that it is not clear what weight to assign to
a given bulk geometry, or even what geometries to include in the sum.

In \cite{farey}, the authors studied the D1-D5 system, and demonstrated that the field theory elliptic genus,
once suitably re-summed, could be interpreted as a sum over the $SL(2,Z)$
images of the BTZ black hole.
One surprising result was that the weight of each geometry was not simply the Euclidean action, but rather had an unexpected (from the bulk point of view) additional factor with a definite modular weight.

We begin our analysis in section 2, by reviewing some basic facts and 
definitions about the BTZ black hole and the problem associated with 
Poincar\'e recurrences.

In section 3, we compute the bulk correlator in the same infinite family of
geometries that appeared in \cite{farey}, and study the sum.
 We show that all but finitely many of the terms receive large
perturbative corrections, absent in the index of ref~\cite{farey}. These
corrections are small in the BTZ background and in finitely many of its images under
$SL(2,Z)$, but they become large for all the other terms.

 Even if the sum (including these large corrections) converges, the result is {\em not} quasi-periodic; it suffers from essentially the same
difficulty as the result computed simply in the BTZ spacetime.
The rest of the paper addresses this question in more detail. In particular,
we discuss the role of an interesting critical time at which {\em all} the
contributions  from different $SL(2,Z)$ images of the BTZ metric
to the correlator are of the same order, despite the fact that they are
exponentially supressed compared to the leading term.

Specifically, in section 4, we exhibit a simple toy model that exhibits part
of the difficulty in recovering a quasi-periodic result from an expansion
around the classical limit.  In this toy model, an exact re-summation of the
perturbation series is required to detect the periodicity in time. In our bulk result, we study the interplay of perturbative and
non-perturbative corrections arising from the saddle point approximation,
and argue that such a re-summation is likely to be extremely difficult--perhaps even ill defined.  We estimate the back-reaction due to the probe operator itself, and show that this interferes with reliably computing the correlator after the same critical time mentioned above, even in the geometries which do not receive large corrections near $t=0$.

In section 5, we discuss how to give a bulk spacetime interpretation to
Poincar\'e recurrences.

In section 6, we discuss some of the implications for the information paradox
and conclude.

Appendices A, B, C and D detail the derivation of two point functions and the action in the BTZ background as well in its $Z_N$ orbifolds.

\section{Poincar\'e Recurrences and Heisenberg Time}

We begin by briefly reviewing the problem.  Consider the Kruskal extension of the AdS-Schwarzschild
black hole.  This geometry has two AdS boundaries, connected only
by spacelike trajectories that pass through the interior of the black hole (see Fig. \ref{adsbh}).  This classical geometry suggests that the Hilbert space of the boundary theory is the tensor product of two of the usual Hilbert spaces, with no interaction term in the Hamiltonian.  Analytic continuation from Euclidean space defines a particular bulk state, the thermal field double or Hartle-Hawking state, which is pure but entangles the two sides, and has the property that the expectation value of any product of operators inserted only on one side is equal to the trace of the operators in the thermal ensemble in the original Hilbert space.  The theory in this state is completely equivalent to thermal field theory in the ordinary thermal ensemble; in particular, correlators involving operators inserted on both boundaries are simply analytic continuations of ordinary thermal correlators.  For a brief review, see \cite{trouble,kos}.

\begin{figure}
\centering \epsfxsize=2.3 in \hspace*{0in}\vspace*{.2in}
\epsffile{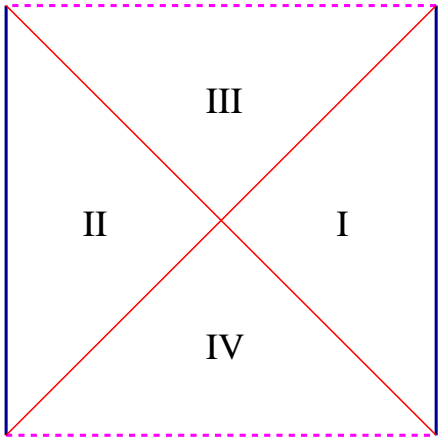}
\caption{The extended conformal diagram of the BTZ black hole.  The regions I and II are outside the black hole, each with its own asymptotic AdS boundary.  Regions III and IV are inside the horizon.  The two boundaries are connected by spacelike trajectories passing through the horizon.}
\label{adsbh}
\end{figure}

In thermal field theory, finite entropy implies that the spectrum of the Hamiltonian is discrete. In such systems, there exist Poincar\'e recurrences--given an initial configuration, because of the finite phase space volume the system
generically evolves under unitary time evolution in such a way that it comes arbitrarily close to its initial state an infinite number of times.  While often discussed in the context of classical physics, this phenomenon extends to the behavior of correlators in quantum theories (see  \cite{dls}, the appendix of \cite{disturb}, and \cite{br} for a discussion in this context and additional references).  In particular, under some weak assumptions about the operator $\phi$, one can prove that the time-like
two point function $\la \phi(t) \phi(0) \ra \equiv f(t) \la \phi(0) \phi(0)
\ra$ is an almost periodic function of time\foot{We use the terms almost periodic and quasi-periodic interchangeably here, with the following definition:  $f(t)$ is almost periodic if $\forall \, \epsilon > 0$, $\exists$ $T$ such that in all intervals $I_T$ of length $T$, $\exists$ a $\tau \, \in \, I_T$ such that $|f(t + \tau) - f(t)| < \epsilon, \, \forall \, t$.}, with an average value $(1/T) \int^T |f(t)|^2 dt \sim \exp(-S)$ , where $S$ is the entropy in the canonical ensemble. The details of the time dependence of $f(t)$ depend very sensitively on the details of the spectrum, but generically the expected time $T$ between order one recurrences is at least exponentially long in the entropy.

On the other hand, correlators computed in black hole spacetimes are exponentially damped in time, and in particular are {\em not} almost periodic.  They always (in the Hartle-Hawking state, at least) take their maximum value at $t=0$, and due to the damping never come back to that value again.  For the same reason, their time-averaged value is zero.  In \cite{malda}, this second problem was solved by adding another geometry to the calculation of the correlator.  In $n+2$ Euclidean dimensions, there is at least one Euclidean geometry with the same asymptotics as the AdS black hole ($S^1 \times S^n$); namely, global AdS with Euclidean time periodically identified.  Continuing to Lorentzian space gives two copies of AdS in the Hartle-Hawking state.  Correlators computed in this geometry will be exactly periodic in time, because of the usual harmonic property of AdS.  If we weight the contribution from this geometry by the exponential of its Euclidean action, we obtain a correlation function with the correct long-time average (see Fig. \ref{corr}).

\begin{figure}
\centering \epsfxsize=6in \hspace*{0in}\vspace*{.2in}
\epsffile{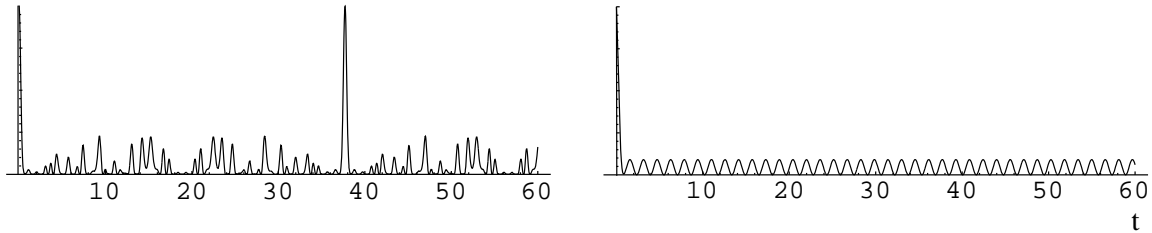}
\caption{On the right, qualitative behavior of $f(t)$ computed by summing the BTZ black hole and thermal AdS, in the high-temperature regime where the black hole dominates.  On the left, a typical quasi-periodic function with the same long time average.}
\label{corr}
\end{figure}

However, the first problem--that the correlator is not almost periodic--remains \cite{br}.  In what follows we compute a more exact correlator by summing over the infinite ensemble of Euclidean geometries with the same asymptotics as the BTZ black hole, but we again fail to obtain a quasi-periodic result.  The reason is essentially that all of these geometries, with the exception of thermal AdS, contain a black hole, and correlators computed in geometries with horizons always exponentially damp to zero in time.  As in \cite{malda}, the contribution of thermal AdS provides a purely oscillating contribution which remains forever, but with an exponentially suppressed amplitude.

\section{Topological Diversity}

For $d=3$  the conformal boundary topology of the AdS black hole is a two dimensional torus, and there are infinitely many smooth, negative curvature Einstein
metrics that can fill a given boundary torus.  The saddle point 
approximation to the correlator is then an infinite sum over these fillings, but as we will show, 
it is not quasi-periodic.

One way to try to fix this is to include bulk spaces with singularities in 
the sum. In particular, we will consider 'mild' singularities (orbifolds). 
These give corrections that can be relevant at large times, 
but again will not reproduce the recurrences.

\subsection{BTZ Black Holes}

We are interested in a 1+1 dimensional thermal field theory on a circle. The thermal condition can be expressed as a periodic compactification of Euclidean time, with period $\beta = 1/T$, and with antiperiodic boundary conditions for fermions. This means we should study a Euclidean field theory on a torus with periodicities $(\phi, t_E) \rightarrow (\phi + 2 \pi n, t_E + \beta m)$, 
$n,m\in Z$, and a flat metric:

\eq
\label{torus}
ds^2 = d\phi^2 + dt_E^2 = (2 \pi)^2 |d\sigma_1 + \tau d\sigma_2|^2 ,
\eeq
where $\sigma_1 = \phi/2 \pi$, $\sigma_2 = t_E/\beta$ and $\tau = i \beta/2 \pi$.
In general $\tau$ can be complex, with ${\rm Im}(\tau) = \beta/2 \pi$ the inverse temperature, 
and ${\rm Re}(\tau)$ related to the chemical potential.  In what follows, we will be interested in the case ${\rm Re}(\tau) = 0$.  The coordinates $\sigma_i$ have periodicity $1$.

\begin{figure}
\centering \epsfxsize=4in \hspace*{0in}\vspace*{.2in}
\epsffile{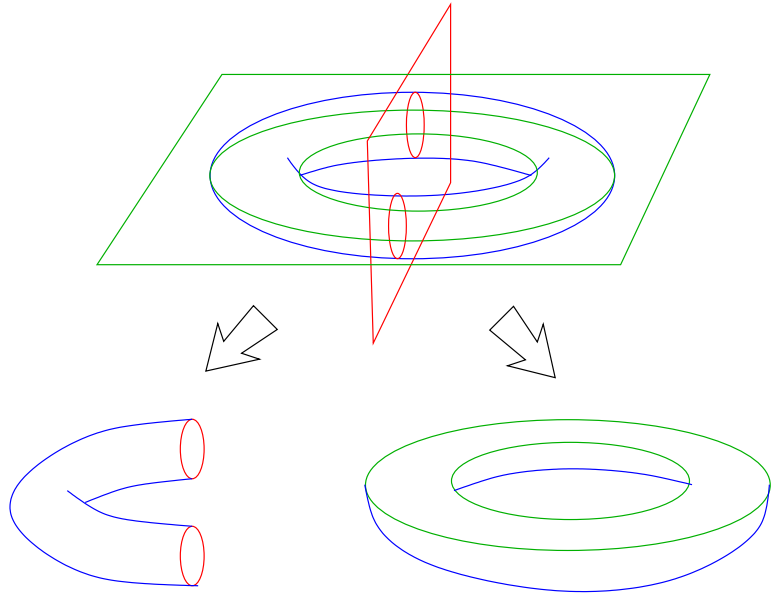}
\caption{\small Two possible slicings of a solid torus as $t=0$ data for Lorenztian space times: on the left the `donut slicing', giving two copies of AdS in an entangled state; on the right the `bagel slicing', giving the BTZ black hole.}
\label{cut}
\end{figure}

According to the prescription discussed above, we should look for three dimensional
Euclidean metrics such that the conformal boundary is a two torus, with a boundary metric in the same conformal class as the one above \cite{mms,farey}.  One class of such examples is certain rotating BTZ black holes:

\eq
ds^2 = \frac{(r^2 - r_+^2)(r^2 - r_-^2)}{l^2 r^2} dt^2 + \frac{r^2 l^2}{(r^2 - r_+^2)(r^2 - r_-^2)} dr^2 + r^2 \left(d\phi + \frac{r_+ (i r_-)}{r^2} dt\right)^2 ,
\label{BTZmetric}
\eeq
where $l$ is the AdS radius (in what follows we set $l=1$). These metrics are parametrized by two real numbers $(r_+, ir_-)$.
They describe solid tori, under the identification
$z\equiv \phi + it  \sim z + 2\pi m + 2\pi \tau' n $, $m, n\in Z$
The conformal boundary is a 2-torus with modular parameter $\tau'=
\tau_1' + i \tau_2'$: $r_+ = \tau_2'/|\tau'|^2$ and 
$r_- = - i \tau_1'/|\tau'|^2$.

To identify the field theory on the torus described by (\ref{torus}) with the conformal boundary of the black hole, we need to map the two up to a Weyl transformation; i.e., find a diffeomorphism which maps $(\sigma_1,\sigma_2)$ of the field theory to $(\sigma_1',\sigma_2')$ of the black hole conformal boundary, such that $ds^2(\tau) = \Omega^2 ds^2(\tau')$. That can be done by an $SL(2,Z)$ transformation:

\begin{eqnarray} 
\left(
\begin{array}{c}
\sigma_1' \\
\sigma_2' 
\end{array}
\right)
& = &
\left(
\begin{array}{cc}
a & -b \\
-c & d
\end{array}
\right)
\left(
\begin{array}{c}
\sigma_1 \\
\sigma_2 
\end{array}
\right), \nonumber \\
\tau' & = & \frac{a \tau + b}{c \tau + d} \, .
\end{eqnarray}

The two point function for a free scalar field of mass $m$,\foot{We measure all masses in units of the AdS radius.} is (see Appendix A):

\eq
\sum_{n \in Z} 
\left |2 (a\tau +b) \sin \left[
\frac{\phi + i t_E +2 \pi n(c\tau + d)}{2(a\tau + b)}\right]\right|^{-4\Delta},
\eeq
where the conformal weight $2 \Delta = 2 \bar{\Delta} = 1 + \sqrt{1 + m^2}$. Notice that the two point function is invariant under the transformations:

\begin{eqnarray}
\label{trans}
\left(
\begin{array}{cc}
a & b\\
c & d
\end{array}
\right)
& \rightarrow  &
\left(
\begin{array}{cc}
a & b \\
c+a  & d+b
\end{array}
\right).
\end{eqnarray}

This invariance is a consequence of the fact that, in a solid torus, there is a unique choice of a contractible 1-cycle $[\gamma_1]$.  However, the non-contractible cycle $[\gamma_2]$ is determined only up to adding multiples of $[\gamma_1]$; that is, $[\gamma_2] \rightarrow [\gamma_2 + n \gamma_1]$ for any integer $n$.  This transformation is generated by eq. (\ref{trans}), which is a $STS$ transformation ($S$: $\tau \rightarrow -1/\tau$, $T$: $\tau \rightarrow \tau + 1$). Therefore to avoid overcounting we should sum only over $SL(2,Z)/Z$, not over the full $SL(2,Z)$ \cite{farey}.  In figure \ref{plane} we have represented the metrics that contribute to our problem. It is convenient to plot the $-1/\tau$ plane, where the $STS$ symmetry is simply a T-translation. These black holes are parametrized by a pair of co-prime integers $(a,b)$.  The integers $(c,d)$ are determined by $(a,b)$ up to an $STS$ transformation.

\begin{figure}
\centering \epsfxsize=3.3in \hspace*{0in}\vspace*{.2in}
\epsffile{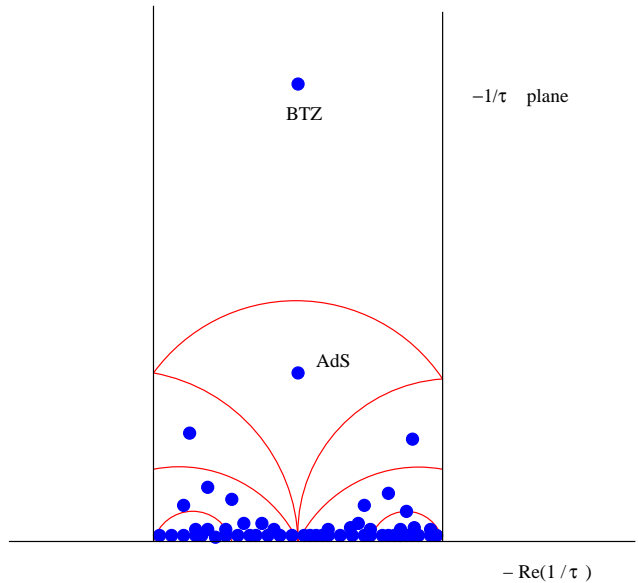}
\caption{\small Black holes contributing to our problem are represented as 
dots in the $-1/\tau$ plane. At high temperature the action
of each of these black holes approaches zero as we come close to the real axis.}
\label{plane}
\end{figure}

Presumably, each of these configurations should be weighted by $e^{-I}$, where $I$ is the Euclidean action. The action is proportional to the volume of the space, which is infinite for spaces with an AdS conformal boundary. However, in the context of AdS/CFT we understand how to regulate this divergence. The regularized action is (see appendix \ref{regularized-action}):

\eq
\label{action}
I = - \pi k r_+ = - \frac{i \pi k}{2} \left( \frac{1}{\tau'} - \frac{1}{\bar{\tau}'} \right) = - \pi k \frac{\tau_2}{|a \tau + b|^2}= - \frac{2 \pi^2 k \beta}{a^2 \beta^2 + 4 \pi^2 b^2},
\eeq
where $k = c/6 = 1/4G_N$ and $c$ is the central charge of the boundary theory. For small temperatures ($\beta > 2 \pi$) the dominant contribution is thermal AdS. For high temperatures  ($\beta < 2 \pi$) the dominant contribution is the nonrotating BTZ.

We can now construct the correlator in this ``ensemble" of geometries, by simply summing the contributions with a weight given by (\ref{action}).  However, there is immediately a problem.  Consider the part of the sum with $a=1$, $b$ large.  From (\ref{action}), $I \rightarrow 0$, so each term has an equal weight.  Therefore the partition function will diverge, as will the correlators.
 Since the action tends to zero in this limit, perturbative corrections can be larger than the classical action.
One such correction has been estimated in~\cite{carlip}. It changes the action
as
\eq
I \rightarrow  I + \delta I, \qquad
\delta I =  -3\log|\tau'|. \label{carl}
\eeq
It is easy to find $SL(2,Z)$ images of the BTZ black hole where the
correction $\delta I$ is at least
as large as the classical action. For instance, take
$a=1, b=n, c=0, d=1$. At high temperature, $\beta \ll 1$, the correction
becomes of order the classical action when $n^2 \sim \beta k$.  For these geometries, $k r_+ \sim 1$,
so the black hole radius is at the Planck scale.
Such geometries will receive large corrections, which we do not know how to estimate.

If the partition function is to converge, these large $b$ terms must be suppressed.  At any rate, we can take this pathology as a signal that whenever subdominant saddles become
important for the computation of the two-point function, then the entire
semiclassical approximation breaks down.

\subsection{Lorentzian Correlators}

We define the Lorentzian two point function by analytically continuing the Euclidean one, setting $i t_E = t$:\footnote{To avoid unnecessary complications with
the continuation, we set $2 \Delta$ equal to a positive integer.}
\eq
e^{\pi k \frac{\tau_2}{|a \tau + b|^2}}
\sum_{n \in Z}
\left(2 |a\tau +b|^2 \sin \left[
\frac{\phi + t +2 \pi n(c\tau + d)}{2(a\tau + b)}\right]
\sin \left[\frac{\phi - t +2 \pi n(c\bar{\tau} + d)}{2(a\bar{\tau} + b)}\right]
\right)^{-2\Delta} .
\eeq
This function has poles on the light cone $\phi \pm t = 2 \pi N$, where $N$ is an integer.\foot{We have ignored the $i \epsilon$ prescription here, as we will focus on the ``two sided" correlator, which has no lightcone singularities.}
These are the usual divergences one expects when the operators are null separated (there is an infinite number of such points, since the theory is on a circle).  They are perfectly physical, but it is more convenient to work with the ``two sided" correlator, obtained by continuing $i t_E = t + i \beta/2$, which
is finite for all times and has no poles:
\eqa
&& e^{\pi k \frac{\tau_2}{|a \tau + b|^2}}
\sum_{n \in Z} \nonumber \\
&&\left(2 |a\tau +b|^2 \sin \left[
\frac{\phi + t + i \beta/2 +2 \pi n(c\tau + d)}{2(a\tau + b)}\right]
\sin \left[
\frac{\phi - t - i \beta/2 +2 \pi n(c\bar{\tau} + d)}{2(a\bar{\tau} +
b)}\right]\right)^{-2\Delta} . \nonumber \\ &&
\eeqa
For large $t$, $a \neq 0$, and fixed $n$ the time dependence is:
\eq
\la O(t+ i \beta/2) O(0) \ra \sim \exp\left( \pi k \frac{\tau_2}{|a \tau + b|^2} \right) \exp\left( - 2 \Delta \frac{|a|\tau_2}{|a \tau + b|^2} t \right) .\label{twopoint}
\label{timedep}
\eeq
(Recall that $\tau_2 = - i \tau = \beta/2 \pi$, for the case of zero chemical potential.)

The first exponential comes from the action. The full result is the sum over all geometries with fixed boundary, and hence over all relatively prime integers $(a,b)$.  The contribution with $a = 0$  is thermal AdS, and is purely oscillatory.  All the other terms are black holes, and have two point functions that decay with time. Notice that this decay can be very slow for configurations with large $b$.
These contributions can be important at very long times because the leading terms decay more quickly, but their amplitude is very small (see Fig. \ref{plots}).

As we mentioned above, perturbative corrections to the classical
action grow out of control for most of these
``spinning'' Euclidean BTZ black holes, so, whenever these configurations
give a significant contribution to the two-point function, the semiclassical
approximation breaks down. In this interpretation, the (perturbative)
bulk computation of the correlator is not reliable past that critical time.

A very interesting fact is that because of a cancellation between the suppression factor from the action and the slower decay rate, the critical time
is the same for all $SL(2,Z)$
images, independently of $a$ and $b$, namely it is $t_0 \sim \pi k/2 \Delta $.  To give an interpretation to this time, note that the amplitude of the leading contribution to the correlator has decayed by a factor of $\exp(-S)$ at $t=t_0$.  In general, we expect finite entropy effects to become important at times later than $t_0$, so it is precisely when finite entropy matters that the calculation becomes unrealiable.  We will discuss this further later in this section and in section 4.  The importance of this time was also noted in \cite{malda,kos}.

\begin{figure}
\epsffile{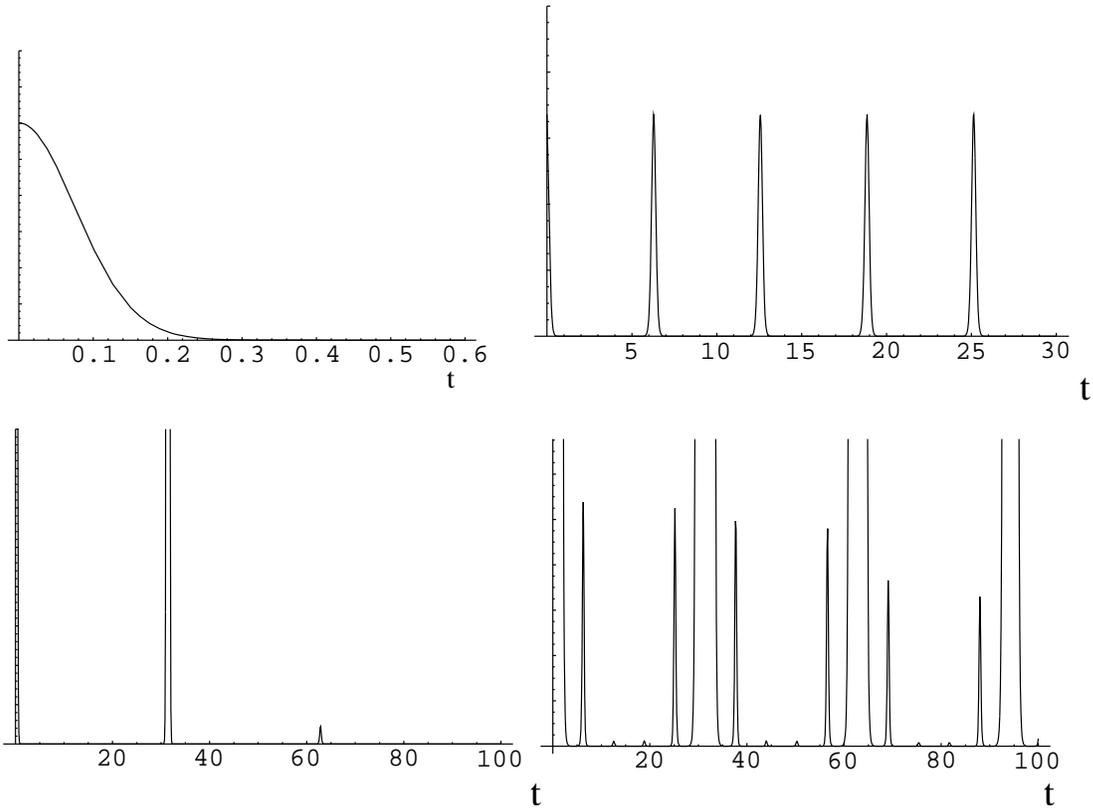}
\caption{\small Two point correlators as functions of Lorentzian time 
$t$ for different geometries. We have taken $\beta = 0.95$ and $k=1$ (larger $k$ would give an exponential weight to the non-rotating BTZ). The upper left figure is the non-rotating BTZ black hole, which is the dominant contribution at high temperature. On the upper right is thermal AdS, which gives an oscillating function with period $2 \pi$ in AdS units. On the lower left is the $a=1$, $b=10$ rotating black hole. There are peaks which become relevant at long times, but the overall contribution is decaying. The lower right figure 
is a detail of the lower left.}
\label{plots}
\end{figure}

\subsection{Orbifold Singularities}

We have seen that the sum over smooth geometries,  even if convergent, gives a decaying
two point function at large $t$,
and is unable to reproduce the expected quasi-periodicity. One possible fix is to add some singular bulk geometries.  Without understanding the structure of the theory better, we do not know which geometries should contribute.
Our only guide is the calculation of \cite{farey}, so we will follow that work and consider Euclidean geometries that have an orbifold singularity at some point in the bulk (but of course with the same boundary torus).~\footnote{
Ref.~\cite{farey} considered a particular index that receives no contribution
from black hole states, so one may question the wisdom of focusing on the
same class of bulk metrics that were considered there. Here, we only
use~\cite{farey} as a justification for including certain singular metrics
that would be discarded in a naive semiclassical approximation to quantum
gravity.}

If we allow a conical deficit angle $2 \pi (1-1/N)$, with $N$ a natural number, the solution can be constructed easily by taking a $Z_N$ identification. In AdS, this is simply the usual way to define a point particle, where the mass is proportional to the deficit angle:
\eq
M = \frac{1}{8 G_3} \left( 1 - \frac{1}{N}\right).
\eeq

In Appendix C we compute the two point function of these Euclidean orbifolds, and in Appendix \ref{regularized-action} we compute their action.  The Euclidean two point function, including the action, is:
\eq
\sum_{n=1}^N\sum_{m\in Z} \kappa \, \exp\left(
\frac{\pi k \tau_2}{|a \tau + b|^2 N^2} \right)
\left| 2 N (a\tau +b) \sin \left[
\frac{u +2 \pi m (c\tau + d)+ 2 \pi n (a \tau + b)}{2 N(a\tau + b)}\right]
\right|^{-4\Delta},
\eeq

For $i t_E = t + i \beta/2$:
\begin{eqnarray}
\la O(t+ i \beta/2) O(0)\ra &  =  &
\sum_{n=1}^N\sum_{m\in Z} \kappa \, \exp\left(
\frac{\pi k \tau_2}{|a \tau + b|^2 N^2} \right) (2N)^{-4\Delta}
|a\tau +b|^{-4\Delta}
 \nonumber \\ && \left( 
\sin \left[\frac{\phi + t + i \beta/2 +2 \pi m (c\tau + d)+ 2 \pi n 
(a \tau + b)}{2 N(a\tau + b)}\right ] \right.\nonumber \\
&&
\left. 
\sin \left[\frac{\phi - t - i \beta/2 +2 \pi m (c\bar{\tau} + d)+ 2 \pi n
(a \bar{\tau} + b)}{2 N(a\bar{\tau} + b)}\right]\right)^{-2\Delta}.
\end{eqnarray}

The long time behavior ($a \neq 0$) of these functions is:
\eq
\langle O(t+ i \beta/2) O(0) \rangle \sim  \exp \left( -
\frac{|a|\tau_2}{|a \tau + b|^2}t \right) . \label{lto}
\eeq

Naively, the correlator decays as
$2\Delta t / N$ rather than $t$ as above, but one can easily check that, thanks to the sum over the $Z_N$
images, all contributions decaying as $\exp[ -(2\Delta+l)
(|a|\tau_2/|a \tau + b|^2)t / N)]$ vanish for all integers $0<l<N-2\Delta$.
Therefore,
$Z_N$ orbifolds contribute terms to the two-point function that
decay in time much more slowly (by a factor of
$2\Delta$) than in a rotating black hole background
[eq.~(\ref{timedep})].  However, the action for them differs by a
factor of $1/N^2$ (see appendix \ref{regularized-action}), so their amplitude
is suppressed.  As in the case of the $SL(2, Z)$ images, this implies that
there is a critical time $t_0$ where the contributions from these are of the
same order as the leading term.  Including the oribifold geometries once again fails to reproduce the correct time dependence.

  Specializing to the case of the non-rotating black hole (
$\tau = i \beta/2\pi$), the terms in the two-point function coming from the
$Z_N$ orbifold geometries will be of order the contribution of the black hole
at a time $t_0$ satisfying
$2 \pi t_0/\beta + I_{BH}/N^2 = 4 \pi \Delta t_0/\beta + I_{BH}$,
which yields (using $I_{BH} = - 2 \pi^2 k / \beta$ and taking $N$,
$\Delta$ large):

\eq
t_0 \sim \pi k / 2 \Delta,
\eeq
which is also the time where the $SL(2, Z)$ images of the black hole make
order one contributions.  So at $t=t_0$, {\em every geometry considered here makes a contribution of the same order to the two point function}.   A crucial point is that since all the terms take their global maximum at $t=0$,
it is impossible for the result to be quasi-periodic, 
even if the corrections are such that the sum converges.

\section{Large Entropy Limit}

The entropy of a conformal field theory is proportional to the central charge.  The limit $c = 6 k \rightarrow \infty$ corresponds to $G_N \sim 1/k \rightarrow 0$. The infinite entropy limit of the field theory is the classical gravity limit of the bulk (and of course is also the limit where the gravitational entropy diverges).  We will argue that in order to recover Poincar\'e recurrences, we need information about the entire expansion in $k^{-1}$. In general the series will contain both perturbative, i.e. $O(1/k^n)$, and non-perturbative, $e^{-\alpha k}$ terms.  In this section we will consider some simple examples that illustrate how large entropy limits behave.

\subsection{A Toy Model}

Consider a particle on a circle of radius $R = k$. The spectrum is discrete, with $\Delta E = 1/k$. In the limit $k \rightarrow \infty$, the spectrum becomes continuous and the entropy infinite. A two point function in the thermal ensemble will be of the form:

\eq
f(t) \equiv \la O(t) O(0) \ra = \sum_n \frac{e^{-\beta E_n}}{Z(\beta)} \la n | O(t) O(0) |n \ra = \sum_{n,m} \frac{e^{-\beta E_n}}{Z(\beta)} e^{i t (E_n -E_m)} |O_{n m}|^2  ,
\eeq
where $O_{n m} \equiv \langle n | O(0) |m \rangle$.
The energies are $E_n = |n|/k$ for every integer $n$, and hence:
\eq
Z(\beta) = \frac{1 + e^{-\beta/k}}{1 - e^{-\beta/k}}.
\eeq
To take a simple example, consider an operator $O$ such that $|O_{n m}|^2 = \delta_{n,0} + \delta_{m,0}$, then
\eq
Z(\beta) f(t-i \beta/2) = 1 + {\sinh(\beta/2k) \over \cosh(\beta/2k) - \cos(t/k)},
\eeq
where we have regulated the divergences by taking $t \rightarrow t - i \beta/2$.
This is a manifestly periodic function of $t$, with period $k$.\footnote{The
function is exactly periodic because of the simple form of the
spectrum; interactions will shift the energy levels and generically make the
time dependence quasi-periodic.}
However, consider the expansion around the continuous spectrum at $k = \infty$:
\eqa
f(t-i \beta/2) &=& {2 \beta^2 \over 4 t^2 + \beta^2} + {\beta \over 2 k} +
{\beta^2 (4 t^2 - 3 \beta^2) \over k^2(96 t^2 + 24 \beta^2)} + 
{\beta^3 \over 24 k^3} + 
 \nonumber\\ && {\beta^2(16 t^4 - 24 t^2 \beta^2 + 25 \beta^4) \over 1920
k^4 ( 4 t^2 + \beta^2)} + {\cal O}(1/k^5).
\eeqa

In the limit of infinite entropy $f(t) \sim 2\beta^2/(4 t^2 + \beta^2)$, and the function is not periodic. To recover the exact periodicity we would need to resum the entire series.  It is
evident that as $t$ increases, one must keep more and more terms in the sum to get an accurate estimate of the value of the function.  For this simple model, the power series is absolutely convergent for any $t$, so this is possible. 
One could determine with greater and greater confidence that the 
function is periodic, by keeping more and more terms in the $1/k$ expansion.  
However, the behavior is quite different in an expansion around a saddle,
where if there are other saddles contributing, the series is asymptotic (and
the divergent part of the series, when appropriately re-summed, gives
information about the sub-leading saddles~\cite{berry}).  In particular, there is a limit to the accuracy that can ever be obtained by expanding only around one saddle.

\subsection{Winding String}

A more interesting example, that has been analyzed in \cite{malda}, is a field $X(\sigma, \tau)$ that lives on a circle of radius $R \sim k$. This field theory is a simple model of the long string model of the black hole of \cite{ms}. In \cite{malda} the two point correlator of the operator 
$O = \sum_n \partial X \bar{\partial} X(\tau, \sigma +2 \pi n)$ was computed:

\beq
\langle O(\phi, t) O(0,0)\rangle \sim \sum_{n,m} \left[
\cosh(2 \pi (t + 2 \pi m k)/\beta) + \cosh(2 \pi (\phi + 2 \pi n)/\beta)
\right]^{-2}.
\eeq

The function is periodic in time with period $k$. For $\phi =0$ and times $|t| < \pi k$ the leading term in the large $k$ limit is:

\beq
\langle O(0, t) O(0,0)\rangle \sim \sum_{n} \left[ \cosh(2 \pi t/\beta) + \cosh(4 \pi^2 n/\beta)\right]^{-2} .
\eeq

The correlation function decays exponentially with time, with a decay rate given by the temperature. The corrections are of the form $e^{-4 \pi^2 |m| k /\beta}$:

\begin{eqnarray}
\delta \langle O(0, t) O(0,0)\rangle  & \sim  &\sum_{n, m > 0}  e^{-8 \pi^2 m k /\beta} \left[ \left( e^{2 \pi t/\beta} +  e^{- 4 \pi^2 m k/\beta}  \cosh(4 \pi^2 n/\beta) \right)^{-2}  \right. \nonumber \\
& & \left. +  \left( e^{-2 \pi t/\beta} +  e^{- 4 \pi^2 m k/\beta}  \cosh(4 \pi^2 n/\beta) \right)^{-2}  \right]  .
\end{eqnarray}

\subsection{Breakdown Time Scale}

We have seen that at the critical time $t_0 \sim \pi k / 2 \Delta$, all the geometries we are considering make a contribution of the same order to the two point function.  Close to $t=0$, on the other hand, the amplitude of the 
black-hole contribution is exponentially larger.  One way to think about the $Z_N$ orbifold geometries, as discussed in \cite{farey}, is that they represent the back-reaction of Hawking particles on the geometry.  The conical defects are interpreted as virtual particles near the horizon.  Keeping for the moment only two geometries, the BTZ black hole  and its $Z_N$ orbifold for $N \gg 1$, we obtain the ratio (for large $t$):

\eq
\left( \la \phi(t) \phi(0) \ra_{BH} + \la \phi(t) \phi(0) \ra_{Z_N} \right) / \la \phi(t) \phi(0) \ra_{BH}  \sim  1 + e^{ \left( 4 \pi \Delta t - 2 \pi^2 k \right) / \beta}.
\eeq
If we can regard the $Z_N$ geometries as representing the effect of perturbations around the black hole, we can see clearly from this equation that the perturbations become large at the critical time $t_0$.\foot{The reader may wonder where there is room in the usual computation of Hawking radiation for such order one effects.  The answer is presumably that the quantity that is being corrected is already exponentially small.}
This gives an indication that we can not trust the semi-classical approximation for $t > t_0$, even in the finite number of geometries where the black hole is larger than string scale.  Since in fact {\em all} the relevant geometries are of the same order at this time, an honest calculation would require taking them all into account, including perturbative corrections around them.  We will discuss this further in the next section.

\section{Beyond Gravity}

There are various ways we could imagine addressing the mystery of the missing recurrences.
The most radical is to rethink the meaning of the AdS/CFT
``duality.'' It could be that only the CFT side of the correspondence
is well-defined as a quantum theory, and that {\em any}
bulk description, including the one given by perturbative, closed string theory
is incomplete. A more modest goal is instead to find a ``phenomenological''
description of Poincar\'e recurrences in AdS gravity language.
Precisely, the question would not be how to predict the Poincar\'e recurrences
in AdS, but rather to ask what the bulk spacetime looks like on time scales of
order the recurrence time.

\subsection{``Phenomenological'' Description of Recurrences}
One option is to
modify the state in which we are computing the correlator.
Continuation from Euclidean space defines a unique Lorentzian correlator, but
if we modify the state in the Lorentzian spacetime away from the Hartle-Hawking state, we can cook up more complicated time dependences.
Normally such modifications lead to divergent terms in the expectation value
of the stress tensor at the horizon, and as such are inconsistent.
However, it is possible to avoid this, as we demonstrate below.

Consider for simplicity a non-rotating BTZ black hole. It has a single
horizon: $r_+=r_H, r_-=0$. A generic scalar
Green function that depends only on $t-t'$ and $\phi-\phi'$ reads:
\eqa
G(t,\phi,r| t',\phi',r')&=&G_{HH}(t-t',\phi-\phi',r,r') + \nonumber \\
&& \sum_m
\int d\omega H_m(\omega) \bar{F}_m(\omega,r) F_m(\omega,r')e^{-i\omega (t-t') +im
(\phi-\phi')}. \label{1}
\eeqa
Here, $F_m(\omega,r)\exp(im\phi -i\omega t)$ denotes a normalizable solution
of the wave equation, with frequency $\omega$ and angular momentum $m$.
The explicit form of the solution can be found for instance in~\cite{kv}.
Near the boundary
it behaves as $ F_m(\omega,r)\approx r^{-2\Delta} $. Near the horizon it 
becomes~\cite{kv,hkv}
\beq
e^{-i\omega t + im\phi} F_m(\omega,r)\approx
e^{-i\omega t +im\phi}\left[e^{i\omega (\beta/4\pi)\log(r-r_H)} + 
e^{2i\theta(\omega) - i\omega (\beta/4\pi)\log(r-r_H)}\right]. \label{2}
\eeq 
The function $\exp[2i\theta(\omega)]$ has poles at
\beq 
\omega = inr_H, \quad n=1,2,3,...; \qquad 
\omega =  \pm m -2inr_H  -2i r_H \Delta, \qquad n=0,1,2,...\; . \label{3}
\eeq
Here, we used eqs.~(3.17,3.18) of ref.~\cite{bl}, and we set $r_-=0$, 
$r_+=r_H$, $\Lambda=1$, $h_+=\Delta$. 
The Green function $G_{HH}$ can be used to compute the expectation value of
the stress-energy tensor on a black-hole geometry, 
$\langle T_{\mu\nu} \rangle$ which is finite on both future and past 
horizons. We want to preserve this property, so we have to make sure that
the additional contribution to $\langle T_{\mu\nu} \rangle$ in eq.~(\ref{1}) is
also well behaved. For the trace of the stress-energy tensor,
the contribution can be written as  
\beq
\langle \delta T_\mu^\mu \rangle \approx 
\sum_m \int d\omega H_m(\omega)\omega^2 {\beta \over 4\pi} 
\left[ e^{2i\theta(\omega)+ (-1 -i\omega/r_H)\log(r-r_H) } + 
c.c.\right].
\eeq
For $r-r_H \ll 1$, the integral of the first term in brackets is evaluated
by closing the contour in the upper half-plane. If the poles of $H_m(\omega)$
satisfy $\Im \omega > r_H$, then, thanks to eq.~(\ref{3}), the integral is
finite. The $c.c.$ term in the integral is evaluated by closing the contour
in the lower half-plane, resulting in the condition $|\Im \omega|>r_H$
[cfr. eq.~(3.19) in~\cite{bl}].

Notice that when $t-t' \gg 1$, the Green function eq.~(\ref{1}) always decays
exponentially. On the other hand, if $H_m(\omega)$ has a pole of order $N+1$
at $\omega_0$, then the Green function behaves as
$(t-t')^N \exp(-\Im\omega_0 |t-t'|)$, so it grows then decays.

Of course, this {\em ad hoc} modification of the Lorentzian Green function
cannot predict the recurrence time, but at best describe it.
At this point it may be useful to draw a parallel with thermodynamics.
Thermodynamics can describe the evolution of macrostates obtained by
averaging over many macroscopically indistinguishable microstates. It can be
used reliably to describe the approach of a macroscopic system to
equilibrium, but not to predict Poincar\'e recurrences. Likewise, the bulk
description of AdS gravity may be able to predict macroscopic properties of
spacetime (such as Hawking radiation), but not its fine details: recurrences,
the encoding of information in Hawking radiation, the discreteness of the
dual CFT spectrum and so on. In this view, the best that we can do is try to
give a ``phenomenological" bulk description of features fully describable only in the dual
CFT language. The {\em ad hoc} modification to the two-point function is just
such an attempt. It shows that a Poincar\'e recurrence can be described as
a small deviation from the thermal vacuum at $t=0$ that evolves in time in such
a way as to produce a resurgence at some later time. In this description,
spacetime is always close to BTZ, since the initial state was chosen
to have negligible backreaction on the metric.

A different yet
related possibility is to allow the backreaction to become big at some
intermediate time. This signals a macroscopic departure of the metric
from the BTZ background. This possibility should not be discarded, since the
recurrence time is much bigger than $t \sim \pi k/ 2 \Delta$--that is, the average
time of a large fluctuation that converts a BTZ black hole into a thermal gas
of light particles.

Finally, one could simply cut off the horizon with a 't Hooftian ``brick wall."  As explained nicely in \cite{br}, such a wall acts as an IR cutoff in the effective Schr\"odinger problem for bulk fields, quantizing their spectrum, making the entropy finite, and hence necessarily leading to a (quasi) periodic result for the bulk two point correlator.  However, this is highly unnatural from the point of view of the Euclidean geometries.  Euclidean thermal AdS and the BTZ black hole have identical topologies, and cutting off the black hole horizon (which is simply the origin in the Euclidean space) would correspond to arbitrarily cutting out a tube in the thermal AdS.

\subsection{Asymptotics}

There is another option, which unfortunately is very difficult to study without a better understanding of the path integral.
The asymptotic expansions of one-dimensional integrals can exhibit a variety of interesting
behaviors, known collectively as Stokes' phenomena.  In particular, as a function of some
parameter in the integrand, the steepest-descent integration contour can change discontinuously,
causing sub-dominant saddle points to appear or disappear (this typically occurs along a co-dimension one line in complex parameter space, known as a Stokes' line).  Further deformation of the parameter can cause a sub-dominant saddle to become dominant (an anti-Stokes' line; see \cite{berry} for a concise and elegant discussion).

There is an instructive example of such a phenomenon in a situation related to the one studied here.  In the AdS black hole in $d \geq 4$, there is a family of spacelike geodesics connecting
points on the two disconnected boundaries, which have the bizarre property that they asymptote to a null geodesic for special pairs of points.  As discussed in \cite{bhsing}, this naively leads to the (impossible) conclusion that the spacelike separated two-point function has a lightcone divergence.
However, after a careful analysis of the Euclidean continuation, this apparently dominant saddle turns out to be off the contour of integration in the relevant region of parameter space (the parameter here is simply Lorentzian time), and therefore does not contribute at all.  The naive saddle point approximation gives a totally wrong result.

This leads to another possibility for recovering the recurrences.  Suppose that not all the various bulk geometries contribute to the calculation of the two-point function for all values of $t$.  For example, for small $t$ only the BTZ might contribute, and then for larger and larger values of $t$ more contributions might appear.  One could cook up a scenario in which the two point function is quasiperiodic, by judiciously adding in more and more geometries at late times.  This would probably require a weight for the individual geometries in the sum so that the partition function is divergent, if summed over all the contributions.  More seriously, it is again completely {\it ad hoc} without a much more detailed understanding of the gravitational path integral.

\section{Conclusions}

In this paper we attempted to reproduce the long time dependence of the two point correlator in a finite entropy thermal field theory by computing bulk correlators in an infinite number of geometries with fixed conformal boundary.
We have shown that the summation over the geometries that appeared in
the calculation of the elliptic genus in~\cite{farey}
suffers from perturbative corrections that grow large for the ``spinning''
Euclidean black holes. These corrections nullify any attempt to use
semiclassical bulk computations to compute Poincar\'e recurrences, which
occur at times exponentially longer that the time $t_0\sim \pi k/2\Delta$,
at which the ``bad'' saddle points become significant.   We have identified two forms of
corrections to the correlator in these geometries--first, that all but finitely many of the $SL(2,Z)$ images contain black holes
smaller than the string/Planck scale, and have leading corrections that render the $SL(2,Z)$ sum divergent, and second, that even in the geometries with large black holes, the back-reaction effects of the probe become of order unity at $t=t_0$.
It is interesting that the time $t_0$ is also the time when an observer would first begin to detect finite entropy effects in the form of exponentially small fluctuations.  Reproducing
the recurrences would probably require an exact re-summation of the
perturbative and non-perturbative contributions to the sum, and perhaps an
understanding of the integration contours and generalized Stokes' phenomena as
well.

To some of us, the most intriguing question which arises in the consideration
of these issues is the following:  if gravity emerges only as a saddle 
point in some path integral, do we expect an 
observer living in a large, weakly curved space containing a black hole to 
measure finite entropy effects in Hawking radiation?  ``Non-thermalities" (that is,
finite entropy effects) in the spectrum of Hawking particles will always be
invisible if one expands only around the leading saddle point.   The sum over geometries may not be enough to restore unitarity and resolve the information paradox, but it is both necessary and predicted by
the AdS/CFT correspondence.

Large black holes in AdS are thermodynamically stable, that is, they have
positive specific heat. However, if their entropy is finite as is expected
from the Gibbons-Hawking calculation, they should undergo Poincar\'e
recurrences.  Such events would be forbidden in classical gravity, because
they will involve a decrease in the horizon area, but will be consistent with
global conservation laws.  One possible (but exponentially unlikely) event is
that the black hole completely dissolves into particles, which in an AdS time form a
black hole again.  Such an intermediate state would look similar to thermal
AdS in the unstable (high temperature) phase.  It is then very tempting to
consider the thermal AdS contribution to the correlator as representing this
process. Another process that would signal departure from perfect
thermal equilibrium was described in section 5: a small deviation
from equilibrium at time $t=0$ focuses at later times to produce a large
fluctuation in the two-point function of a certain observable.
Both effects could be measured quite easily by an observer in the AdS Schwarzschild space, but it will require a much better
understanding of the theory before we can address these issues definitively.

\subsection*{Acknowledgments}
We would like to thank D. Birmingham, J. Maldacena, G. Moore, E. Rabinovici, 
N. Seiberg, S. Shenker, L. Susskind, and E. Verlinde for helpful discussions. 
The work of M.K. is supported by NSF Grant PHY-0332258. M.P. is supported in 
part by NSF grant PHY-0245068. R.R. is supported by DOE under grant 
DE-FG02-90ER40542.

\appendix

\section{Two Point Function in the BTZ Metric}
In the usual metric for the hyperbolic plane
\eq
ds^2= {dz^2  + dw d\bar{w} \over z^2}, \label{hyp}
\eeq
one gets the correlation function for a field of conformal weight $(\Delta, \bar{\Delta})$:
\eq
\langle O(w) O(w')\rangle = \kappa \, (w-w')^{-2 \Delta}(\bar{w}-\bar{w}')^{-2 \bar{\Delta}} ,
\eeq
where $\kappa$ is a constant that it is irrelevant to our discussion.

Here we will consider operators associated to scalar particles in the bulk,
i.e. $2\Delta = 2\bar{\Delta} = 1 + \sqrt{1 + (m l)^2}$, where $m$ is the mass
of the particle and $l$ is the AdS radius (from now on, we will set $l = 1$). 
For particles of mass $m \gg 1$, $2\Delta \sim m  + 1 + O(1/m)$, i.e. 
the mass of the bulk particle and the conformal weight of the dual operator 
are approximately equal.

To avoid unnecessary complications with the analytic continuation to Lorentzian
signature we set $2\Delta$ equal to a natural number.

The two-point function in the black hole geometry is
written in terms of the torus coordinate
$u = (\phi + i t_E)$, related to $w$ by $w=\exp [iu/(a \tau + b)]$.

Under a change of coordinate $w=f(u)$, a conformal correlator changes as
\eq
\langle O_{new}(u) O_{new}(u')\rangle =
\left| {f'(u)f'(u')\over (f(u) -f(u'))^2}\right|^{2\Delta}.
\eeq
In our case $f'(u) = i(a\tau + b)^{-1} f(u)$; so the
correlator is:
\eq
\left|2 (a\tau +b) \sin \left[\frac{u +2 \pi n(c\tau + d)}{2(a\tau + b)}\right]
\right|^{-4\Delta}.
\eeq

In terms of the coordinate $w$, the identification that gives rise to the black 
hole metric is\footnote{In the usual BTZ $(r, \phi, t_E)$ coordinates:
\begin{eqnarray}
\nonumber
w & = & \left( \frac{r^2 - r^2_+}{r^2 - r^2_-} \right)^{1/2}
e^{ i(\phi + i t_E)/(a \tau + b)}, \nonumber \\
z & = & \left( \frac{r^2_+ - r^2_-}{r^2 - r^2_-} \right)^{1/2} e^{ r_+ \phi + ir_- t_E},
\nonumber
\end{eqnarray}
where $r_+ = \tau_2/|\tau|^2$ and  $r_- = - i \tau_1/|\tau|^2$.}
\eq
w\sim w \exp(2 \pi i/\tau') ,
\eeq
where $\tau' = (a \tau + b)/(c \tau + d)$ and for our problem $\tau=i\beta/2\pi$.
Then the sum over images gives:
\eq
\sum_{a,b,c,d \in \Gamma_\infty\/ \Gamma, n \in Z} \kappa \, \exp(-I -\delta I)
\left|
2 (a\tau +b) \sin \left[\frac{u +2 \pi n(c\tau + d)}{2(a\tau + b)}\right]
\right|^{-4\Delta},
\eeq
where
$I$ is the classical action:
\eq
I = - \frac{i \pi k}{2} \left( \frac{1}{\tau'} - \frac{1}{\bar{\tau}'} \right)
= - \pi k \frac{\tau_2}{|a \tau + b|^2},
\eeq
 and $\delta I$ represents peturbative corrections, for instance as in
eq.~(\ref{carl}).

\vspace{2cm}

\section{$SL(2,C)$}

The infinitesimal isometries of the metric eq.~(\ref{hyp}) are:
\eq
w\rightarrow \alpha + \beta w + \gamma w^2 -\bar{\gamma}z^2,
\qquad
z\rightarrow (\Re\beta + \gamma w + \bar{\gamma}\bar{w})z, 
\qquad \alpha,\beta,\gamma \in C.
\eeq
They realize an $SL(2,C)$ that takes the boundary $z=0$ into itself, and
there it acts on $w$ in the usual manner:
\eq
w\rightarrow {Aw + B\over Cw + D} ,\qquad A,B,C,D \in C, \qquad AD-BC =1.
\eeq
Up to conjugations in $SL(2,C)$, any element of $SL(2,C)$ can be reduced
to one of the following two forms:
\eq
\left(\begin{array}{cc} \exp\theta & 0 \\ 0 & \exp(-\theta) \end{array}\right),
\qquad
 \left(\begin{array}{cc} 1 & 1 \\ 0 & 1 \end{array}\right). \label{10}
\eeq
In particular, the first form gives both thermal AdS ($\theta=\beta$)
and the BTZ black hole ($\theta = 4\pi^2/\beta$).

Now choose two commuting elements in $SL(2,C)$,
$T_1, T_2$, and consider the discrete group $G=\{nT_1 + mT_2, m,n\in Z\}$.
We want to sum over all geometries that are asymptotically a torus, so the
general question is, when is $C/G$ a torus? Up to now, we considered only one 
case, that gives both thermal AdS and the Euclidean BTZ black hole:
\eq
T_1= \left(\begin{array}{cc} \exp(2\pi i/\tau) & 0 \\ 0 & \exp(-2\pi i/\tau)
\end{array} \right),
\qquad T_2= \left(\begin{array}{cc} 1 & 0 \\ 0 & 1 \end{array}\right).
\eeq
We can also consider the torus generated by two non-collinear translations
in the $(w,\bar{w})$ plane. They give the extremal black hole at nonzero
temperature. This configuration is singular, but our attitude is that we should
allow for such configurations; judiciously. However, it is not difficult to see that none of the tori generated by two translations give any significant
contribution to the long-time behavior of the two-point function.

The most general group $G$ is obtained as follows. Using a conjugation in
$SL(2,C)$, $T_1$ can be cast in the form given in eq.~(\ref{10}). Now, write
$T_2$ as
\eq
\left(\begin{array}{cc} A & B \\ C & D \end{array}\right),
\eeq
and impose that it commutes with $T_1$. This gives us the following cases:
\begin{enumerate}
\item When $T_1$ is the identity, $T_2$ is
arbitrary, but we can still use conjugations in $SL(2,C)$ (they commute with
$T_1$ of course!) to recast it in either diagonal or upper-triangular form.
In the first case, if $\theta\neq 0$ we get the identifications of BTZ or
AdS type. If $\theta=0$, $C/G$ is not a torus. When $T_2$ is upper-triangular,
we do not get a torus from the quotient.
\item When $T_1$ is upper triangular, $T_2$ is also upper triangular of the
form:
\beq
\left(\begin{array}{cc} 1 & B \\ 0 & 1 \end{array}\right).
\eeq
Iff $\Im B\neq 0$ we get a torus.
\item When $T_1$ is diagonal with $\theta\neq 0$, then $T_2$ is also diagonal
\eq
\left(\begin{array}{cc} \exp(\vartheta) & 0 \\ 0 & \exp(-\vartheta)
\end{array}\right).
\eeq
When $\theta=\vartheta$ we get a torus, because the angular coordinate in
$z$ is automatically compact. When $\theta\neq\vartheta$ we must have
either $\theta=2\pi i k/N$ or $\vartheta=2\pi i k/N$ to get a torus. This is
the point-particle case and its $SL(2, Z)$ images.
\end{enumerate}

\vspace{2cm}

\section{$Z_N$ Orbifolds and Two Point Functions}

Here we will compute the two point functions for $Z_N$ orbifolds of the BTZ black hole. As BTZ is itself a $Z$ orbifold of global AdS, we can construct these spaces with a hyperbolic quotient (the BTZ quotient or thermal AdS) and an elliptic quotient of finite order.

Let us take the point particle in thermal AdS, and then construct all the others by $SL(2,Z)$ transformation.
We consider the two identifications:
\eq
w\sim \exp(2\pi i/N)w, \qquad w \sim \exp(\beta/N) w = \exp(-2\pi i \tau/N)w.
\eeq

The first identification yields the $Z_N$ orbifold in the angular coordinate and the second one the finite temperature identification of thermal AdS. To get an angular coordinate with the usual periodicity we have to take 
$w = e^{iu/N}$, $u=\phi + i t_E$.

Using the standard rules for transforming conformal fields as in appendix A, and
summing over images, we get
\eq
\langle O_{new}(u) O_{new}(u')\rangle=
\kappa \, \sum_{n=1}^N\sum_{m\in Z}
\left|2 N \sin \left({u + 2\pi (n-m\tau) \over 2N}\right)\right|^{-4\Delta}.
\eeq
The Lorentzian continuation is $it_E \rightarrow t + \pi \tau$; then the
two-point function at $\phi=0$ becomes
\eqa
\langle O_{new}(t) O_{new}(0)\rangle&=& 
\kappa \, \sum_{n=1}^N\sum_{m\in Z+1/2}\nonumber \\ &&
\left|(2 N)^2 \sin \left({t + 2\pi  (n-m\tau) \over 2N} \right)
\sin \left({-t + 2\pi  (n-m\bar{\tau})\over 2N}\right)\right|^{-2\Delta}. 
\nonumber \\ &&
\eeqa
Notice that for $n=0$, $m=1/2$, $t=2\pi N$,
and $N\gg 1$ the two-point function is
$O(1)$
\eq
\langle O_{new}(2\pi N) O_{new}(0)\rangle\approx
\left({2\over \beta} \right)^{2\Delta}. \label{orbi}
\eeq

The generalization to the BTZ black holes is straightforwardly obtained by
applying $SL(2,Z)$ transformations to eq.~(\ref{orbi}), in complete analogy
with the BTZ case worked out in Appendix A.
In Euclidean time the two point function is thus:

\eq
\sum_{n=1}^N\sum_{m\in Z} \kappa \, \exp(-I)
\left|2 N (a\tau +b) \sin \left[
\frac{u +2 \pi m (c\tau + d)+ 2 \pi n (a \tau + b)}{2 N(a\tau + b)}\right]
\right|^{-4\Delta},
\eeq
It can be analytically continued to Lorentzian signature by the same 
substitution we used earlier: $it_E\rightarrow t +\pi \tau$.

\vspace{2cm}
\section{Regularized Action for the BTZ Black Holes}
\label{regularized-action}

In this appendix we evaluate the regularized action for the Euclidean BTZ black hole by using the method of counterterms \cite{hkl,bk,ejm}. This method is based on the fact that all the divergences that appear in the Einstein-Hilbert action, plus the Gibbons-Hawking term, can be cancelled by a finite set of boundary integrals. In three dimensions we have to sum three contributions:
\begin{eqnarray}
I_{\cal M} & = & - \frac{1}{16 \pi G_N}\int_{\cal M} \sqrt{g}  (R + 2/l^2), \\
I_{\partial{\cal M}} & = & - \frac{1}{8 \pi G_N} \int_{\partial{\cal M}} 
\sqrt{h} K, \\
I_{c} & = & \frac{1}{8 \pi G_N} \int_{\partial{\cal M}} \sqrt{h}  ,
\end{eqnarray} 
where $g_{ij}$ is the 3 dimensional BTZ metric, $l$ is the AdS radius, K is the trace of the extrinsic curvature, and $h_{ij}$ is the induced metric. The first contribution is the usual Einstein-Hilbert action for AdS spaces, regulated by putting a cut-off surface $\partial{\cal M}$ at the radius $r=l/\epsilon$. The second contribution is the Gibbons-Hawking boundary term at the cut-off, and the last term is the counterterm.

By explicit evaluation with the metric eq. (\ref{BTZmetric}) we obtain:
\begin{eqnarray}
I_{\cal M} & = & \frac{\beta}{4 G_N} \frac{1}{\epsilon^2} 
\left(1-\epsilon^2 \frac{r_+^2}{l^2} \right), \\
I_{\partial{\cal M}} & = & - \frac{\beta}{4 G_N}\frac{1}{\epsilon^2} 
\left[ 1 + \left(1 - \epsilon^2  \frac{r_+^2}{l^2}\right) 
\left(1 - \epsilon^2 \frac{r_-^2}{l^2}\right) \right], \\
I_{c} & = & \frac{\beta}{4 G_N} \frac{1}{\epsilon^2} 
\sqrt{\left(1 - \epsilon^2 \frac{r_+^2}{l^2}\right) \left(1 - 
\epsilon^2 \frac{r_-^2}{l^2}\right)}.
\end{eqnarray}

Summing the three contributions and taking the $\epsilon \rightarrow 0$ limit we find:

\beq
I_0 =  \frac{\beta}{8 G_N} \frac{r_-^2 -r_+^2}{l^2}.
\eeq

Using the relation between the temperature, $r_+$, and $r_-$, we have:

\beq
I_0 = - \frac{\pi r_+}{4 G_N} = -\pi k \frac{r_+}{l} .
\eeq

\vspace{0.5cm}
If we are interested in a Euclidean non-rotating BTZ black hole with a conical deficit at the horizon then we have to include a contribution from the curvature singularity (see \cite{su}):

\beq
I_{sing}  =  - \frac{1}{16 \pi G_N}\int_{sing} \sqrt{g} R =  - \frac{r_+}{4 G_N} \delta ,
\eeq

where $\delta$ is the deficit angle $\delta = 2 \pi - \beta r_+/l^2$. The total gravitational action is then:

\beq
I_g \equiv I_{sing} + I_0 =  \frac{1}{8 G_N} \frac{\beta r_+^2}{l^2} - 
\frac{\pi r_+}{2 G_N}.
\eeq

It is interesting to write this action in terms of $\beta$ and the deficit angle:
\beq
I_g =  - \frac{\pi^2}{2 \beta G_N} \left[ 1 - (\delta/2\pi)^2 \right].
\eeq

If we interpret this conical deficit as created by a Euclidean 
``point particle'' of mass $m$ we have to include its action:
\beq
I_{pp} = m L =  \frac{\pi^2}{\beta G_N} \frac{\delta}{2 \pi}\left[ 1 - (\delta/2\pi) \right] .
\eeq

And then the total action is:
\beq
I_{tot} = I_g + I_{pp} =  - \frac{\pi^2}{2 \beta G_N} \left[ 1 - (\delta/2\pi) \right]^2 =  -\frac{\beta}{8 G_N} \frac{ r_+^2}{l^2}.
\eeq

\end{document}